\documentclass[11pt,eqs]{article}
\usepackage{latexsym}
\usepackage{amsmath}
\usepackage{hyperref}
\hypersetup{
	colorlinks=true,
	linkcolor=cyan,
	filecolor=magenta,      
	urlcolor=cyan,
}

\makeatletter
\newcommand\xleftrightarrow[2][]{%
	\ext@arrow 9999{\longleftrightarrowfill@}{#1}{#2}}
\newcommand\longleftrightarrowfill@{%
	\arrowfill@\leftarrow\relbar\rightarrow}
\makeatother

\title{
	Double space T-dualization and coordinate dependent RR field
	\thanks{Work supported in part by the Serbian Ministry of Education,Science and Technological Development.}}
\author{B. Nikoli\'c and D. Obri\'c \thanks{email:dobric, bnikolic@ipb.ac.rs}\\{\it Institute of Physics Belgrade, University of Belgrade, Pregrevica 118, Serbia} }

\begin{document}
	
	\maketitle
	
	\begin{abstract}
	In this article we examine T-dualization in double space formalism of type II superstring theory in pure spinor formulation. Background fields that we consider will all be constant except Ramond-Ramond field which will infinitesimally depend only on bosonic coordinates $x^\mu$. In double space T-dual transformations are represented as permutations between starting $x^\mu$ and dual coordinates $y_\mu$. Combining these two sets of coordinates into double coordinate $Z^M = (x^\mu, y_\mu)$ while demanding that T-dual double coordinates have same transformation laws, we obtain how background fields transform under T-duality. Comparing these results with ones obtained with Buscher T-dualization procedure we are able to conclude that these two approaches are equivalent for cases where background fields have coordinate dependence. 
	\end{abstract}

\section{Introduction}
\setcounter{equation}{0}

One of most peculiar properties of superstring theory is presence of net that connects all different types of superstrings \cite{Zwiebach,Polchinski2,Polchinski,Becker Becker Schwarz}. Spanned by two dualities, called T-duality and S-duality, this network hints at existence of a much grander theory lurking underneath it, M-theory. While it is not yet exactly known what shape M-theory will take, it is certain that by exploring properties of string theory dualities we will be able to gain some insight into this underlying theory. In this paper we will focus on exploring properties of only T-duality.

T-duality is a phenomenon which does not appear in  any shape in theories that deal with one dimenional particles. This phenomenon is only experienced by extended objects, strings \cite{Polchinski,Becker Becker Schwarz,T-duality procedure 1,Rocek Verlinde,T-duality explained 2,Alvarez Alvarez-Gaume Barbon Lozano} and it connects theories which have radii of compactification $R$ with ones where radii is $\alpha \slash R$. Basic mathematical framework for implementing T-duality is given with Buscher procedure \cite{T-duality procedure 1,T-duality procedure 2} which relies on existence of underlying global isometries in theory under consideration. In short, in order to obtain T-dual theory, we localize global isometry, usually translational isometry, by introducing covariant derivatives. Covariant derivatives inadvertently introduce new gauge fields into the theory. In order for the number of degrees of freedom to stay the same between starting theory and its dual we need to eliminate all newly introduced degrees of freedom with Lagrange multipliers. By using gauge freedom we are able to fix starting coordinates, which leaves us with theory only dependent on gauge fields and Lagrange multipliers. By finding equations of motion for gauge fields and inserting their solutions back into the theory, we obtain T-dual theory. While this procedure is applicable for many models it breaks down when we have coordinate dependent background fields. In cases where background fields only infinitesimally depend on coordinates, it is possible to generalize Buscher procedure \cite{Lj. B. S. T-dualizacija 1,Generalized Buscher procedure,Davidovic Nikolic Sazdovic,ljdbs} by introducing invariant coordinates given as line integral of covariant derivatives.

Even though Buscher T-dualization procedure can be considered as definition of T-duality, in order to gain deeper understanding of duality, it is useful to consider alternative formulations. One of these alternative representations, called "double space" formulation, casts T-duality as permutation of coordinates in space spanned by coordinates from both staring and T-dual theories. Double space formulation was first considered in papers \cite{oDD 1,dd1,dd2,dd3,dd4}, recently this formalism has been related with $O(D,D)$ transformations \cite{oDD 2,dd5,dd6,dd7,dd8}. While in papers \cite{oDD 1,dd9,dd10,dd11,dspace 1,dspace 2,dd12,dd13} T-dualization along some directions has been represented as permutation of these coordinates with T-dual ones.

In articles \cite{dspace 1,dspace 2} it was shown that for type II superstring theory, Buscher procedure and double space formalism  are equivalent. Analysis was conducted on type II superstring in pure spinor formulation where all background fields were held constant. Here we would like to repeat procedure outlined in aforementioned articles while allowing Ramond-Ramond field to be dependent on bosonic coordinates.

In this article we will begin by showing how we obtained our action and background fields from general one. By integrating out fermionic momenta we will obtain theory which has R-R field coupled with derivatives of bosonic coordinates. If this step was omitted then theory would have R-R field coupled only with fermionic degrees of freedom which will not be dualized here. After obtaining starting action, we will also give presentation of results that are obtained with Buscher T-dualization procedure. By transcribing T-dual transformation laws in term of double coordinates, generalized metric and generalized current, while demanding that T-duality does not alter the form of transformation laws, we will be able to obtain T-dual generalized metric and T-dual generalized current. By equating components of starting and dual generalized metric and generalized current we are able to show how background fields transform under T-duality.  Comparing these results with ones obtained from Buscher procedure will will be able to see if these two approaches differ for cases with coordinate dependent background fields.

\section{Type II superstring theory with coordinate dependent RR field and its dual}
\setcounter{equation}{0}
In this section our goal is to introduce action for type II superstring in pure spinor formulation \cite{pure spinor formalism papers 1, pure spinor formalism papers 2, pure spinor formalism papers 3, pure spinor formalism papers 4}. We will be working with theory that has all background fields, except Ramond-Ramond field, constant. Ramond-Ramond field will depend only on bosonic coordinates and only infinitesimally, furthermore we will also demand that RR field is antisymmetric. Both constraints are necessary if we wish to obtain "reasonable" transformation laws between starting and T-dual theory.

After introduction of type II superstring, we will present theory which is its T-dual. This theory will have new background fields and it has been show that this theory is both non-commutative and non-associative \cite{Nikolic Obric 1, Nikolic Obric 2, Nikolic Obric 3}.

At the end of this section we will transcribe both theories in notation that is more suitable for double space formulation.

\subsection{Type II superstring in pure spinor formulation} 

Action that describes propagation of type II superstring, in its most general form \cite{Vertex operators}, is given as

\begin{equation}\label{eq:dejstvo}
S=S_0 + V_{SG},
\end{equation}

where first term denotes action for string propagating in constant fields
\begin{equation} \label{S_0 action}
S_0 = \int_{\Sigma}d^2 \xi \left( \frac{k}{2} \eta_{ \mu \nu } \partial_m x^\mu \partial_n x^\nu \eta^{m n} - \pi_\alpha \partial_- \theta^\alpha + \partial_+ \bar{\theta}^\alpha \bar{\pi}_\alpha \right) 
+S_\lambda +S_{ \bar{\lambda} }.
\end{equation}
Here integration is done on a world-sheet $\Sigma$, parametrized by coordinates $\xi^m$, where parameter $m$ takes values $m = 0,1\  (\xi^0 = \tau,\ \xi^1 = \sigma)$. We will work in light-cone coordinates which are given by $\xi^\pm = \frac{1}{2}(\tau \pm \sigma)$ while light-cone partial derivatives are given by $\partial_\pm = \partial_\tau \pm \partial_\sigma$. Superspace is spanned by $10$ bosonic coordinates $x^\mu,\ \mu= 0,1,...,9$ and fermionic ones $\theta^\alpha$ and $\bar{\theta}^\alpha$, with $16$ independent real parameters each  $(\alpha=1,2,...,16)$. Variables $\pi_\alpha$ and $\bar{\pi}_\alpha$ denote canonically conjugated momenta of fermionic coordinates. Terms $S_\lambda $ and $S_{\bar{\lambda}}$ denote actions for pure spinors, $\lambda^\alpha$ and $\bar{\lambda}\alpha$, and their momenta, $\omega_\alpha$ and $\bar{\omega}_\alpha$, where pure spinors satisfy pure spinor constraints
\begin{equation}
\lambda^\alpha (\Gamma^\mu)_{\alpha \beta} \lambda^\beta =  \bar{\lambda} (\Gamma^\mu)_{\alpha \beta} \bar{\lambda}^\beta = 0 .
\end{equation} 
Second term in equation (\ref{eq:dejstvo}) denotes all perturbations to the flat background. These perturbations are given by integrated vertex operator for massless type II supergravity
\begin{equation}
V_{SG} =  \int_{\Sigma}d^2 \xi (X^T)^M A_{MN} \bar{X}^N.
\end{equation}
While in general case matrix $A_{MN}$ contains fields that are dependent on both bosonic and fermionic coordinates. We will be working with following background fields
\begin{equation}
	A_{MN} = \begin{bmatrix}
0 & 0 & 0 & 0 \\
0 & k(\frac{1}{2}g_{\mu \nu} + B_{\mu \nu}) & \bar{\Psi}_\mu^\beta & 0\\
0 & -\Psi_\nu^\alpha & \frac{2}{k}(f^{\alpha\beta} + C_\rho^{\alpha\beta} x^\rho)& 0\\
0 & 0 & 0 & 0
\end{bmatrix},
\end{equation}
where  $g_{\mu \nu}$ is symmetric tensor, $B_{\mu \nu}$ is Kalb-Ramon antisymmetric tensor, $\Psi_\mu^\alpha$ and $\bar{\Psi}_\mu^\alpha$ are Mayorana-Weyl gravitino fields and $\frac{2}{k}(f^{\alpha\beta} + C_\rho^{\alpha\beta} x^\rho) = \frac{2}{k}F^{\alpha\beta}$ is Ramond-Ramond field. Ramond-Ramond field is composed of constant antisymmetric tensors  $f^{\alpha \beta}$ and $C_\rho^{\alpha \beta}$, where $C_\rho^{\alpha \beta}$ is also infinitesimal. We chose to work with antisymmetric $F^{\alpha\beta}$ tensor in order to obtain transformation laws that can be easily recast in double space formulation. Dilaton field $\Phi$ is assumed to be constant, where factor $e^\Phi$ has been incorporated into $f^{\alpha \beta}$ and $C_\rho^{\alpha \beta}$. Since we are only interested in classical analysis, we will not calculate dilaton shift under T-duality transformations. This choice of fields is accompanied with following constrain
\begin{align} \label{constraint}
\gamma^\mu_{\alpha \beta} C_\mu^{\beta \gamma} = 0, \quad \gamma^\mu_{\alpha \beta} C_\mu^{\gamma \beta} = 0.
\end{align}

In general, vectors $X^M$ and $\bar{X}^M$ are given as columns containing partial derivatives of both fermionic and bosonic coordinates as well as containing fermionic momenta and pure spinor contribution. In order to simplify calculations we will neglect all terms that are non-linear in fermionic coordinates $\theta^\alpha$ and $\bar{\theta}^\alpha$. This means vectors $X^M$ and $\bar{X}^M$ have following form
\begin{equation}
	X^M = \left(\begin{matrix}
\partial_+ \theta^\alpha\\
\partial_+ x^\mu\\
\pi_\alpha\\
\frac{1}{2} N_+^{\mu \nu}
\end{matrix} \right), 
\  \bar{X}^M = \left( \begin{matrix}
\partial_- \bar{\theta}^\lambda\\
\partial_- x^\mu\\
\bar{\pi}_\lambda\\
\frac{1}{2} \bar{N}_-^{\mu \nu}
\end{matrix} \right).
\end{equation}
Pure spinor contribution are given as 
\begin{equation}
N_+^{\mu \nu} = \frac{1}{2} \omega_\alpha {( \Gamma^{[\mu\nu]}
	)^\alpha}_\beta \lambda ^\beta, \quad \bar{N}_-^{\mu\nu} = \frac{1}{2} \bar{\omega}_\alpha {(  \Gamma^{ [\mu\nu ]  }     )^\alpha}_\beta \bar{\lambda}^\beta.
\end{equation}
From this point on, since pure spinors are decoupled from the rest of the action, we will be omitting them.

With these assumptions we have that action (\ref{eq:dejstvo}) takes the following form
\begin{align} \label{Action final S}
S       =    
k    \int_{\Sigma}    d^2    \xi      
\left[     \Pi_{ +\mu \nu }   \partial_+     x^\mu    \partial_-   x^\nu
+     \frac{1}{2}      (  \partial_+  \bar{ \theta }^\alpha  +   \partial_+   x^\mu   \bar{ \Psi }_\mu^\alpha     )
\left( F^{-1}  (x)   \right)_{ \alpha\beta }
(    \partial_-    \theta^\beta      + \Psi_\nu^\beta      \partial_-  x^\nu         )
\right] ,  
\end{align}
where we have integrated out fermionic momenta $\pi_\alpha$ and $\bar{\pi}_\alpha$ as well as introduced following tensors
\begin{equation}
\Pi_{ \pm \mu \nu } = B_{\mu \nu} \pm \frac{1}{2} G_{\mu \nu},
\end{equation}
\begin{equation}
F^{\alpha\beta}(x)=f^{\alpha\beta}+C_\mu^{\alpha\beta} x^\mu\, ,\quad (F^{-1}(x))_{\alpha\beta}=( f^{ -1 } )_{ \alpha\beta }  - ( f^{ -1 } )_{ \alpha\alpha_1 }   C_\rho^{\alpha_1 \beta_1}    x^\rho   ( f^{ -1 } )_{ \beta_1\beta }\, .
\end{equation}
Tensor $(F^{-1}(x))_{\alpha\beta}$ has inherited both properties of Ramond-Ramond tensor, that is this new tensor is anisymmetric and coordinate dependence is only tied to infinitesimal tensor.

\subsection{T-dual theory}

Procedure for obtaining T-dual theory for model that has been presented above is described in great detail in papers \cite{Nikolic Obric 1,Nikolic Obric 2,Nikolic Obric 3}. Here we will only mention final results and give brief description of dualization procedure.

Procedure for obtaining T-dual theory is based on extension of standard Buscher procedure \cite{T-duality procedure 1,T-duality procedure 2,Generalized Buscher procedure}. This procedure entails that dualization is to be carried out only along directions of translational isometry. Since we have that Ramond-Ramond field is antisymmetric, we have that action (\ref{Action final S}) is invariant to translations along bosonic coordinates. Next step in procedure is to localize this symmetry, this is accomplished by substituting partial derivatives with covariant ones. Because we are working with theory that has coordinate dependent background fields, it is also required to introduce invariant coordinates. These coordinates are given as line integrals of covariant derivative. These substitutions bring into play new gauge fields which add new degrees of freedom. In order for T-dual theory to maintain same informational content as starting theory, we must remove these new degrees of freedom by introducing Lagrange multipliers. Utilizing gauge freedom we can also fix starting bosonic coordinates which leaves us with theory that is described by only gauge fields and Lagrange multipliers. Finding equations for gauge field and inserting them into the gauge fixed action we are left with T-dual theory.
Implementing Buscher procedure we obtain following T-dual action

\begin{align}\label{bosonic T dual action}
\begin{gathered}
{}^b S = \frac{k}{2} \int_{\Sigma} d^2\xi \Big[ \frac{1}{2} \bar{\Theta}^{\mu\nu}_- \partial_+y_\mu \partial_- y_\nu +\partial_+ \bar{ \theta }^\alpha \left( {}^b F^{-1} \text{\small $( V^{(0)} )$}    \right)_{ \alpha\beta } \partial_- \theta^\beta
\Big.\\
\begin{aligned}
\Big. +\partial_+y_\mu {}^b \bar{ \Psi }^{\mu \alpha}  \left( {}^b F^{-1}  \text{\small $( V^{(0)} )$}    \right)_{ \alpha\beta } \partial_- \theta^\beta
+ \partial_+ \bar{ \theta }^\alpha \left( {}^b F^{-1}  \text{\small $( V^{(0)} )$}    \right)_{ \alpha\beta } {}^b \Psi^{\nu \beta} \partial_- y_\nu \Big].
\end{aligned}
\end{gathered}
\end{align}
Here, $y_\mu$ is a dual coordinate, left superscript ${}^b$ denotes bosonic T-duality and $V^0$ represents following integral

\begin{align}
\begin{gathered}\label{dv0}
\Delta V^{(0) \rho}  =\\   
=\frac{1}{2} \int_{P} d \xi^+ \breve{\Theta}_-^{\rho_1 \rho} \left[ \partial_+ y_{\rho_1}     -  \partial_+ \bar{ \theta } ^\alpha ( f^{- 1 } )_{ \alpha\beta } \Psi_{\rho_1}^\beta
\right]
- \frac{1}{2} \int_{P} d \xi^- \breve{\Theta}_-^{\rho \rho_1} 
\left[    \partial_- y_{\rho_1}  +  \bar{ \Psi }_{\rho_1}^\alpha ( f^{- 1 } )_{ \alpha\beta } \partial_- \theta^\beta
\right] .
\end{gathered}
\end{align}
T-dual tensors that appear in action have following interpretation: $\bar{\Theta}_-^{\mu \nu}$ is inverse tensor of $\bar{\Pi}_{\pm\mu\nu} = \Pi_{ \pm\mu \nu } + \frac{1}{2} \bar{ \Psi }_\mu^\alpha \left( F^{-1}  ( x)   \right)_{ \alpha\beta }\Psi_\nu^\beta = \breve{\Pi}_{\pm\mu \nu } - \frac{1}{2}  \bar{ \Psi }_\mu^\alpha ( f^{ -1 } )_{ \alpha\alpha_1 }   C_\rho^{\alpha_1 \beta_1}    x^\rho   ( f^{ -1 } )_{ \beta_1\beta } \Psi_\nu^\beta $, defined as

\begin{equation}
\bar{\Theta}_\mp^{\mu\nu} \bar{\Pi}_{\pm\nu\rho} = \delta^\mu_\rho,
\end{equation}
where
\begin{gather}
\bar{\Theta}_\mp^{\mu\nu} = \breve{\Theta}_\mp^{\mu \nu}  + \frac{1}{2} \breve{\Theta}_\mp^{\mu \mu_1} \bar{ \Psi }_{\mu_1}^\alpha (f^{-1})_{\alpha \alpha_1} C^{\alpha_1 \beta_1}_\rho V^{(0)\rho} (f^{-1})_{\beta_1\beta} \Psi_{\nu_1}^{\beta}  \breve{\Theta}_\mp^{\nu_1 \nu} ,\\
\breve{\Theta}_\mp^{\mu \nu} \breve{\Pi}_{\pm\nu \rho} = \delta^\mu_\rho,\qquad \breve{\Theta}_\mp^{\mu\nu} = \Theta_\mp^{\mu\nu} - \frac{1}{2} \Theta_\mp^{\mu\mu_1} \bar{ \Psi }_{\mu_1}^\alpha (\bar{f}^{-1})_{\alpha\beta} \Psi^\beta_{\nu_1} \Theta_\mp^{\nu_1\nu}\\
\bar{f}^{\alpha\beta} = f^{\alpha \beta} + \frac{1}{2} \Psi_\mu^\alpha \Theta_-^{\mu\nu} \bar{ \Psi }_\nu^\beta,\\
\Theta_\mp^{\mu\nu} \Pi_{ \pm\mu \rho } = \delta^\mu_\rho, \qquad \Theta_\mp = -4 (G_E^{-1} \Pi_\mp G^{-1})^{\mu\nu},\\
G_{E \mu\nu} = G_{\mu\nu} - 4 (BG^{-1}B)_{\mu\nu},\\
\Pi_{ +\mu \nu } = - \Pi_{ -\nu \mu }, \quad \breve{\Pi}_{+\mu\nu} = - \breve{\Pi}_{-\nu\mu}, \quad \bar{\Pi}_{+\mu\nu} = - \bar{\Pi}_{-\nu\mu},\\
\Theta_+^{\mu\nu} = -\Theta_-^{\nu\mu}, \quad \breve{\Theta}_+^{\mu\nu} = -\breve{\Theta}_-^{\nu\mu}, \quad \bar{\Theta}_+^{\mu\nu} = -\bar{\Theta}_-^{\nu\mu}. 
\end{gather}

Tensor $\left( {}^b F^{-1}  \text{\small $( V^{(0)} )$}    \right)_{ \alpha\beta }$ is T-dual to $\left( F^{-1}  ( x)   \right)_{ \alpha\beta }$, and it is given as
\begin{equation}
\left( {}^b F^{-1} \text{\small $( V^{(0)} )$}    \right)_{ \alpha\beta } = \left( F^{-1}  \text{\small $( V^{(0)} )$}   \right)_{ \alpha\beta } - \frac{1}{2} \left( F^{-1} \text{\small $( V^{(0)} )$}   \right)_{ \alpha\alpha_1 } \Psi_\mu^{\alpha_1} \bar{\Theta}_-^{\mu \nu} \bar{ \Psi }_\nu^{\beta_1} \left( F^{-1}  \text{\small $( V^{(0)} )$}    \right)_{ \beta_1\beta }.
\end{equation}
Finally, $ {}^b \bar{ \Psi }^{\mu \alpha} $ and ${}^b \Psi^{\nu \beta}$ are T-dual gravitino fields, given as
\begin{equation}
\label{Background fields transformation 2}
\quad{}^b \bar{ \Psi }^{\mu \alpha} = \frac{1}{2} \Theta_-^{\mu\nu} \bar{ \Psi }_\mu^\alpha,\quad
\quad{}^b \Psi^{\nu \beta} =  - \frac{1}{2}\Psi_\mu^\beta \Theta_-^{\mu\nu}.
\end{equation}  
Transformation laws that connect starting and T-dual coordinates are 
\begin{align}
\bar{\Pi}_{+\mu\nu} \partial_- x^\nu &= -\frac{1}{2} \partial_- y_\mu - \frac{1}{2} \bar{ \Psi }^\alpha_\mu \left( F^{-1}  (x)   \right)_{ \alpha\beta} \partial_- \theta^\beta - \beta^+_\mu(x), \\
\bar{\Pi}_{+ \mu \nu} \partial_+ x^\nu &= \frac{1}{2} \partial_+ y_\nu - \frac{1}{2} \partial_+ \bar{ \theta }^\alpha \left( F^{-1}  (x)   \right)_{ \alpha\beta} \Psi^\beta_\nu - \beta^-_\nu (x),
\end{align}
where $\beta^+$ and $\beta^-$ functions, which are obtained during T-dualization procedure from varying gauge fixed action with respect to gauge fields, are given as
\begin{align}
\beta^+_\mu(x) &= -\frac{1}{2} ( \bar{ \theta }^\alpha + x^{\nu_1} \bar{ \Psi }^\alpha_{\nu_1}  ) ( f^{ -1 } )_{ \alpha\alpha_1 }   C_\mu^{\alpha_1 \beta_1}    ( f^{ -1 } )_{ \beta_1\beta }  ( \partial_-\theta^\beta + \partial_-x^{\nu_2} \Psi^\beta_{\nu_2} ),\\
\beta^-_\mu(x) &= -\frac{1}{2}  ( \partial_+\bar{ \theta }^\alpha + \partial_+x^{\nu_1} \bar{ \Psi }^\alpha_{\nu_1}  ) ( f^{ -1 } )_{ \alpha\alpha_1 }   C_\mu^{\alpha_1 \beta_1}    ( f^{ -1 } )_{ \beta_1\beta }  ( \theta^\beta + x^{\nu_2} \Psi^\beta_{\nu_2} ).
\end{align}
Having introduced starting and T-dual theory we can now focus on combining them with double space formualism.

\section{T-dualization in double space}\label{sec 3}
\setcounter{equation}{0}
Focus of this section is to show how bosonic T-duality of superstring with coordinate dependent Ramond-Ramond field can be represented as permutation of coordinates in space that is spanned by both starting and dual coordinates. Work done here mirrors work done in papers \cite{dspace 1,dspace 2}, where same model was examined only with constant background fields. While our transformation laws contain both bosonic and fermionic coordinates, it will be possible to separate fermionic contributions into objects called "double currents".

\subsection{T-dual transformation laws in double space formulation}
In order to show how permutations of coordinates can be interpreted as T-dual transformations, we need to transcribe T-dual transformation laws introducing more suitable notation. We begin by introducing following substitutions

\begin{gather}
\Psi^\alpha_\mu = \Psi^\alpha_{+\mu}, \quad \bar{ \Psi }^\alpha_\mu = \Psi^\alpha_{-\mu}, \quad \theta^\alpha = \theta^\alpha_+, \quad \bar{ \theta }^\alpha = \theta^\alpha_-,\\
\left( F^{-1}  (x)   \right)_{ \alpha\beta} = \left( F^{-1}_+  (x)   \right)_{ \alpha\beta}, \quad \left( F^{-1}  (x)   \right)_{ \beta\alpha} = \left( F^{-1}_-  (x)   \right)_{ \alpha\beta},\\
\left( F^{-1}_+  (x)   \right)_{ \alpha\beta} = - \left( F^{-1}_-  (x)   \right)_{ \alpha\beta},\\
( f^{ -1 } )_{ \alpha\alpha_1 }   C_\mu^{\alpha_1 \beta_1}    ( f^{ -1 } )_{ \beta_1\beta } = C_{+\mu\alpha\beta},\quad
( f^{ -1 } )_{ \beta\beta_1 }   C_\mu^{\beta_1 \alpha_1}    ( f^{ -1 } )_{ \alpha_1\alpha } = C_{-\mu\alpha\beta},\\
C_{+\mu\alpha\beta} = - C_{-\mu\alpha\beta}.
\end{gather}

With this new notation, transformation law and its inverse one take the following form

\begin{align}
\partial_\mp x^\nu &= - \frac{1}{2} \widehat{\Theta}^{\nu\mu}_\mp \partial_\mp y_\mu - \frac{1}{2} \widehat{\Theta}^{\nu\mu}_\mp  \left[    \Psi^\alpha_{\mp\mu}  \left( F^{-1}_\pm  \text{\small $( V^{(0)} )$}   \right)_{ \alpha\beta }  - (\theta^\alpha_\mp +  V^{(0)\nu_1} \Psi^\alpha_{\mp\nu_1}   )  C_{\pm\mu\alpha\beta}    \right] \partial_\mp \theta^\beta_\pm, \\
\partial_\mp y_\mu &= - 2 \widehat{\Pi}_{\pm\mu\nu} \partial_\mp x^\nu -  \left[    \Psi^\alpha_{\mp\mu}  \left( F^{-1}_\pm  (x)   \right)_{ \alpha\beta }  - (\theta^\alpha_\mp +  x^{\nu_1} \Psi^\alpha_{\mp\nu_1}   )  C_{\pm\mu\alpha\beta}    \right] \partial_\mp \theta^\beta_\pm , 
\end{align}
where $\beta^\pm_\mu$ functions have been expanded and separated into two parts, one containing partial derivative of bosonic coordinate and other containing partial derivative of fermionic coordinate. First part has been incorporated into two newly introduced tensors $\widehat{\Pi}_{\pm\mu\nu}$ and $\widehat{\Theta}^{\nu\mu}_\mp$
\begin{gather}
\widehat{\Pi}_{\pm \mu\nu} = \bar{\Pi}_{\pm \mu\nu} - \frac{1}{2} (\theta^\alpha_\mp + x^{\nu_1} \Psi^\alpha_{\mp \nu_1} ) C_{\pm\mu\alpha\beta } \Psi^\beta_{\pm\nu},\\
\widehat{\Theta}^{\nu\mu}_\mp =  \bar{\Theta}^{\nu\mu_1}_\mp \left[ \delta^\mu_{\mu_1} + \frac{1}{2}  ( \theta^\alpha_\mp + V^{(0)\nu_1} \Psi^\alpha_{\mp\nu_1}    ) C_{\pm\mu_1\alpha\beta} \Psi^\beta_{\pm\nu_2} \breve{\Theta}^{\nu_2\mu}_\mp    \right].
\end{gather}

These two newly introduced tensor are inverse to one another, $\widehat{\Pi}_{\pm \mu\nu} \widehat{\Theta}^{\nu\rho}_\mp = \delta^\rho_\mu$ and they can be decomposed in following way
\begin{gather}
\widehat{\Pi}_{\pm\mu\nu} = \widehat{B}_{\mu\nu} \pm \widehat{G}_{\mu\nu}, \quad \widehat{\Theta}^{\mu\nu}_\mp = -4 (\widehat{G}_E^{-1} \widehat{\Pi}_\mp \widehat{G}^{-1} )^{\mu\nu},\\
\widehat{G}_{E\mu\nu} = \widehat{G}_{\mu\nu} - 4 \widehat{B}_{\mu\mu_1} \widehat{G}^{\mu_1\nu_1} \widehat{B}_{\nu_1\nu},\\
\widehat{\Theta}^{\nu\mu}_\pm = - 4 \widehat{G}_E^{\nu\nu_1} \widehat{B}_{\nu_1\mu_1} \widehat{G}^{\mu_1\mu} \mp 2 (\widehat{G}^{-1}_E)^{\nu\mu}.
\end{gather}

Interpretation of these components is the following: $\widehat{B}_{\mu\nu}$ and $\widehat{G}_{\mu\nu}$ are antisymmetric and symmetric parts of tensor $\widehat{\Pi}_{\pm\mu\nu}$ called "improved Kalb-Ramond field"  and "improved metric tensor" respectively. Both improved tensors now have additional bilinear form in NS-R fields $\Psi_\mu^\alpha$ and $\bar{ \Psi }^\alpha_\mu$. Tensor $\widehat{G}_{E\mu\nu}$ is called "improved effective metric". These decompositions allow us to rewrite transformation laws as
\begin{gather}
\pm \partial_\pm x^\mu = (\widehat{G}^{-1})^{\mu\nu}\partial_\pm y_\nu + 2 (\widehat{G}^{-1})^{\mu\nu_1} \widehat{B}_{\nu_1\nu} \partial_\pm x^\nu \nonumber\\+ (\widehat{G}^{-1})^{\mu\nu} \left[    \Psi^\alpha_{\pm\nu}  \left( F^{-1}_\mp  (x)   \right)_{ \alpha\beta }  - (\theta^\alpha_\pm +  x^{\nu_1} \Psi^\alpha_{\pm\nu_1}   )  C_{\mp\nu\alpha\beta}    \right] \partial_\pm \theta^\beta_\mp, \label{traa1}\\
\pm \partial_\pm y_\mu = \widehat{G}_{E\mu\nu} \partial_\pm x^\nu - 2 \widehat{B}_{\mu\mu_1} \widehat{G}^{\mu_1\nu}\partial_\pm y_\nu  \nonumber \\ + \frac{1}{2} \widehat{G}_{E\mu\nu} \widehat{\Theta}^{\nu\mu_1}_\pm \left[  \Psi^\alpha_{\pm\mu}  \left( F^{-1}_\mp  \text{\small $( V^{(0)} )$}   \right)_{ \alpha\beta } - ( \theta^\alpha_\pm +  V^{(0)\nu_1} \Psi^\alpha_{\pm\nu_1}    ) C_{\mp\mu_1\alpha\beta} \right] \partial_\pm \theta^\beta_\mp. \label{traa2}
\end{gather} 
It should be noted that all tensors in equation (\ref{traa1}) are functions of bosonic coordinates $x^\mu$. On the other hand all tensors in (\ref{traa2}) are functions of line integral $V^{(0)}$ which has been defined in (\ref{dv0}).

Having transcribed transformation laws in new notation, we are now free to introduce double coordinates 
\begin{equation}
Z^\mu = 	\begin{pmatrix} x^\mu \\ 
y_\mu	
\end{pmatrix}.
\end{equation}
By utilizing double coordinates, transformation laws take surprisingly simple form
\begin{equation} \label{dstl}
\pm \Omega_{MN} \partial_\pm Z^N = \breve{\mathcal{H}}_{MN} \partial_\pm Z^N + \breve{J}_{\pm M},
\end{equation}
where generalized metric is given as
\begin{equation}
\breve{\mathcal{H}}_{MN} = \begin{pmatrix}
\widehat{G}_{E\mu\nu} (V) & -2 \widehat{B}_{\mu\mu_1} (\widehat{G}^{-1})^{\mu_1\nu} (V) \\
2 (\widehat{G}^{-1})^{\mu\nu_1} \widehat{B}_{\nu_1\nu}(x) & (\widehat{G}^{-1})^{\mu\nu} (x)
\end{pmatrix},
\end{equation}
and double current is
\begin{equation}
\breve{J}_{\pm M} = \begin{pmatrix}
\frac{1}{2} \widehat{G}_{\mu\nu_1} \widehat{\Theta}^{\nu_1\nu}_\pm (V)\\
(\widehat{G}^{-1})^{\mu\nu} (x)
\end{pmatrix} J_{\pm \nu}, \quad J_{\pm \nu} = \left[    \Psi^\alpha_{\pm\nu}  \left( F^{-1}_\mp  (x)   \right)_{ \alpha\beta }  - (\theta^\alpha_\pm +  x^{\nu_1} \Psi^\alpha_{\pm\nu_1}   )  C_{\mp\nu\alpha\beta}    \right] \partial_\pm \theta^\beta_\mp. 
\end{equation}
It should be noted that upper components all depend on variable $V^{(0)}$, while lower components all depend on $x$.

The matrix
\begin{equation}
	\Omega_{MN} = \begin{pmatrix}
0 & I_D \\
I_D & 0 
\end{pmatrix},
\end{equation} 
is invariant $SO(D,D)$ metric where $I_D$ denotes unity matrix in $D$ dimensions.

Since generalized metric contains both improved Kalb-Ramond field and improved metric tensor it does not come in standard form. However, it still satisfies following relations
\begin{equation}
\breve{\mathcal{H}}^T \Omega \breve{\mathcal{H}} = \Omega, \quad (\Omega\breve{\mathcal{H}})^2 = I, \quad \Omega^2 = I, \quad \det(\breve{\mathcal{H}}) = 1,
\end{equation}  
which means that $\breve{\mathcal{H}}\in SO(D,D)$ \cite{oDD 1,oDD 2}.

\subsection{T-duality in double space}

To obtain full T-duality let us introduce matrix which permutes starting and dual coordinates
\begin{equation}
{T^M}_N = \begin{pmatrix}
0 & I_D \\
I_D & 0 
\end{pmatrix}.
\end{equation}
Now we can define dual double coordinate as
\begin{equation}
{}^b Z^M = {T^M}_N Z^N.
\end{equation}
Transformation laws (\ref{dstl}) must have the same form for both the dual and starting coordinates, that is
\begin{equation}
\pm \Omega_{MN} \partial_\pm {}^b Z^N = {}^b \breve{\mathcal{H}}_{MN} \partial_\pm {}^b Z^N + {}^b J_{\pm M}.
\end{equation}
From this we can deduce how both generalized metric and double current transform under permutations
\begin{equation}
{}^b \breve{\mathcal{H}}_{MN} = {T_M}^P \breve{\mathcal{H}}_{PQ} {T^Q}_N, \quad {}^b \breve{J}_{\pm M} = {T_M}^N \breve{J}_{\pm N}.
\end{equation}
Expanding first equation we get
\begin{align}
{}^b \breve{\mathcal{H}}_{MN}  &= \begin{pmatrix}
{}^b \widehat{G}_E^{\mu\nu} (V) & -2\  {}^b  \widehat{B}^{\mu\mu_1} \ ({}^b\widehat{G}^{-1})_{\mu_1\nu} (V) \\
2\   ({}^b\widehat{G}^{-1})_{\mu\nu_1} {}^b\widehat{B}^{\nu_1\nu}(x) &  ({}^b\widehat{G}^{-1})_{\mu\nu} (x)
\end{pmatrix} \\
&= \begin{pmatrix}
(\widehat{G}^{-1})^{\mu\nu} (x) &  2   (\widehat{G}^{-1})^{\mu\nu_1} \widehat{B}_{\nu_1\nu}(x)\\
-2   \widehat{B}_{\mu\mu_1} (\widehat{G}^{-1})^{\mu_1\nu} (V) & \widehat{G}_{E\mu\nu} (V)
\end{pmatrix}.
\end{align}
From this we can see that variables $V^{(0)}$ and $x$ also exchange places.

Equating components (2, 2) and (2, 1) we obtain following equations
\begin{equation}
({}^b\widehat{G}^{-1})_{\mu\nu} (x) = \widehat{G}_{E\mu\nu} (V), \quad \rightarrow \quad {}^b\widehat{G}_{\mu\nu} (x) = (\widehat{G}^{-1}_E)_{\mu\nu} (V),
\end{equation}
\begin{equation}
({}^b\widehat{G}^{-1})_{\mu\nu_1} {}^b\widehat{B}^{\nu_1\nu}(x) = -   \widehat{B}_{\mu\mu_1} (\widehat{G}^{-1})^{\mu_1\nu} (V) , \quad \rightarrow \quad {}^b\widehat{B}^{\mu\nu}(x) = -(\widehat{G}^{-1}_E)^{\mu\nu_1}  \widehat{B}_{\nu_1\mu_1} (\widehat{G}^{-1})^{\mu_1\nu} (V).
\end{equation}
Using these two results we are able to obtain 
\begin{equation}
{}^b\widehat{\Pi}_{\pm\mu\nu} (x) =  {}^b\widehat{B}_{\mu\nu}(x) \pm \frac{1}{2}{}^b\widehat{G}_{\mu\nu} (x) = \frac{1}{4} \widehat{\Theta}_{\mp\mu\nu}.
\end{equation}
This result coincides with result obtained from Buscher procedure. 
Equating other two components of the matrix we obtain the same information.

Expanding equation for dual current we obtain
\begin{equation}
{}^b \breve{J}_{\pm M} =
\begin{pmatrix}
\frac{1}{2} {}^b\widehat{G}_{\mu\nu_1} {}^b\widehat{\Theta}^{\nu_1\nu}_\pm (V)\\
({}^b\widehat{G}^{-1})^{\mu\nu} (x)
\end{pmatrix} J_{\pm \nu} = \begin{pmatrix}
(\widehat{G}^{-1})^{\mu\nu} (x)\\
\frac{1}{2} \widehat{G}_{\mu\nu_1} \widehat{\Theta}^{\nu_1\nu}_\pm (V)
\end{pmatrix}J_{\pm \nu},
\end{equation}
where T-dual current has the same factor $J_{\pm \nu}$ while vector components are switched.

Comparing our results to ones obtained in paper \cite{dspace 2}, we notice that while generalized metric, double current and double space transformation laws have same form. However, all fields that emerge in formalism are now modified. These modifications all stem from the fact that starting theory had coordinate dependent RR field. 

\section{Conclusion}
\setcounter{equation}{0}
Aim of this article was to investigate alternative method for obtaining T-dual theories, namely double space method. This investigation was carried out on type II superstring theory with specific choice of background fields. Our choice of model was motivated by the fact that this exact model and its T-dual was examined in great detail in papers \cite{Nikolic Obric 1,Nikolic Obric 2,Nikolic Obric 3}, where T-duality was obtained with Buscher procedure. This work provided us with the reference point on which we can compare our results. 

At the beginning of our examination we described how we obtained our action from more general one. This was done by demanding that all background fields, except Ramond-Ramond field, are constant. Ramond-Ramond field was chosen to have infinitesimal linear dependence on bosonic coordinates $x^\mu$. Furthermore, we also demanded that RR field be totally antisymmetric. Terms that were non-linear in fermionic coordinates have been neglected and all fermionic momenta have been integrated out of the action. These assumptions were necessary in order to obtain simple transformation laws between starting and dual coordinates (for case where RR field is not antisymmetric see \cite{Nikolic Obric 2}). After this we presented T-dual theory, T-dual fields and T-dual transformation laws which were obtained with Buscher procedure. While starting theory is assumed to be commutative, T-dual theory exhibits both non-commutative and non-associative properties. These properties are a consequence of $\beta^\pm_\mu$ functions that show up in transformation laws.

Section \ref{sec 3} was dedicated to double space formulation of T-duality. We began by transcribing T-dual transformation laws and background fields in more suitable notation. By splitting and recasting old background fields into "improved" fields, we were able to gain more clearer picture of underlying connections between starting and dual coordinates. Combining space of starting theory, spanned by $x^\mu$, with space of T-dual theory, spanned with $y_\mu$, we obtain double space formulation.   Double space is now spanned by coordinates $Z^M = (x^\mu, y_\mu)$.  Rewriting T-dual transformation laws into double space coordinates we define two new objects, generalized metric $\breve{\mathcal{H}}_{MN}$ and double current $\breve{J}_{\pm M}$. Their components are expressed through improved Kalb-Ramond field, improved metric tensor and improved effective metric, containing additional terms bilinear in NS-R background $\Psi^\alpha_\mu$ and $\bar{ \Psi }^\alpha_\mu$. Furthermore, it should be noted that these components are not constant, upper row depends on coordinate $x^\mu$ and lower row on its T-dual image $V^\mu$.   

In context of double space formalism, T-duality is given by simple permutation of coordinates. By demanding that double space coordinates $Z^M$ transform as ${}^b Z^M = {T^M}_N Z^N$ and that T-dual coordinates posses same transformation laws as initial ones, we are able to find T-dual generalized metric ${}^b\breve{\mathcal{H}}_{MN}$ and T-dual double current ${}^b\breve{J}_{\pm M}$. Since T-dual generalized metric has the same form as the starting one, by comparing components we are able to deduce expressions for T-dual background fields as functions of starting fields. It should also be noted that permutation of coordinates also permutes arguments of background fields. This means that T-dual generalized metric and T-dual double current now have upper row that depends on $V^\mu$ and lower row that depends on $x^\mu$. 
Doing same analysis for T-dual double current, we obtain relations that connect its components to background fields of starting theory.

Comparing results obtained for T-dual fields by means of double space coordinate permutation with ones obtained with Buscher procedure, it is evident that both methods produce same result. Additionally, it should be noted that comparing results from this paper with results from \cite{dspace 2}, where all background fields were constant, we notice that double space transformation laws have the same form but individual components of generalized metric and double current are now coordinate dependents.




\section{Acknowledgments}

The authors acknowledge that this research has been supported by the Science Fund of the Republic of Serbia, grant 7745968, ``Quantum Gravity from Higher Gauge Theory 2021'' --- QGHG-2021. The contents of this publication are the sole responsibility of the authors and can in no way be taken to reflect the views of the Science Fund of the Republic of Serbia.

\end{document}